\documentclass[onecollarge]{svjour2}
\usepackage{graphicx}
\usepackage{cite}
\usepackage{color}
\usepackage{amsfonts}
\usepackage{soul}
\usepackage{cancel}
\usepackage{amsmath}

\usepackage{amssymb}
\usepackage{esint}

\begin{document}

\title{A formal\footnote{
``I would like to offer some remarks about the word ``formal''.
 For the mathematician, it usually means ``according to the standard of
 formal rigor, of formal logic''.
 For the physicists, it is more or less synonymous with ``heuristic'' as
 opposed to ``rigorous''''.
 Pierre Cartier.
 Mathemagics (A Tribute to L.
 Euler and R.
 Feynman).
 Seminaire Lotharingien de Combinatoire 44 (2000), Article B44d}
 view  on level 2.5 large deviations and fluctuation relations}

\author{Andre C Barato \and Raphael Chetrite}

\institute{
A. C. Barato 
\at II. Institut f\"ur Theoretische Physik, Universit\"at Stuttgart - Stuttgart 70550, Germany\\
\email{barato@theo2.physik.uni-stuttgart.de}\\
\and
R. Chetrite
\at  Laboratoire J. A. Dieudonn\'e, UMR CNRS 6621, Universit\'e de Nice Sophia-Antipolis, France\\
\email{raphael.chetrite@unice.fr}\\
}

\date{Received: date / Accepted: date}

\maketitle
\begin{abstract}
We obtain the rate function for the level 2.5 of large deviations for pure jump and diffusion processes.
This result is proved by two methods: tilting, for which a tilted process with an appropriate typical behavior is considered,
and a spectral method, for which the scaled cumulant generating function is used. We also briefly discuss fluctuation relations, pointing out 
their connection with large deviations at the level 2.5.
\end{abstract}

\section{Introduction}

An important recent progress in nonequilibrium statistical physics
was the discovery of various fluctuation relations, which are identities involving the statistics of a fluctuating entropy.
In particular, the Gallavotti-Cohen-Evans-Morriss (GCEM) relation \cite{evans93,EvansS,Gal} imposes
a peculiar symmetry related to the rare events associated with this fluctuating entropy. The appropriate theory to describe such rare
events is large deviation theory, which is a very fashionable subject in statistical physics \cite{Oono,Touchette} and in modern probability \cite{DenH,Dembo,Deuschel,ellis85,Varadhan},
as evidenced by the Abel Prize awarded to S.R.S Varadhan in 2007. 

We recall that a time dependent measure $\mu_{T}(dx)$ satisfies the large deviation principle if at large times it takes an exponential decreasing form. 
This exponential decay is characterized by a lower semi-continuous positive function $I(x)$, which is called the rate
function. This function is such that for any set $A$ 
\begin{equation}
-\inf_{x\in A^{0}}I(x)\leq\liminf_{T\rightarrow+\infty}\frac{1}{T}\ln\mu_{T}(A)\leq\limsup_{T\rightarrow+\infty}\frac{1}{T}\ln\mu_{T}(A)\leq-\inf_{x\in\overline{A}}I(x),\label{eq:Ldev}
\end{equation}
where $A^{0}$ is the interior of $A$ and $\overline{A}$ is the closure of $A$. This can be stated less formally as
\begin{equation}
\mu_{T}(dx)\sim\exp\left(-TI(x)\right)dx.\label{eq:Ldevf}
\end{equation}

Historically, large deviation theory originated in the nineteenth
century with pioneering works in statistical mechanics \cite{Boltzmann}. One of the
most important contributions to large deviation theory was the general
approach for Markov processes developed by Donsker and Varadhan \cite{Donsker}.
In this series of papers, they identified three levels of large deviations: 
\begin{itemize}
\item Level 1, which is the study of fluctuations of additive observables with respect to the mean. 
\item Level 2, related to fluctuations of the fraction of time spent in each state. 
\item Level 3, concerning fluctuations on the statistics of infinite trajectories. 
\end{itemize}
The ranking of these levels establishes a hierarchy in which a lower level can be deduced from a
higher one by contraction. Donsker and Varadhan proved
the large deviation principle for Markov processes at the level 3
by studying random probability measures on infinite trajectories.
This queen large deviation result posses an explicit rate function,
which is the relative entropy density. Moreover, they
proved the large deviation principle at the level 2 for the empirical
density, defined as the fraction of time spent in each state up to
time $T$. Contrary to level 3, the rate function for level 2 admits
a variational representation, which is in general not explicit. Hence, the explicit character of level
3 disappears after contracting to level 2.
At discrete time a more detailed picture is available: it is possible
to investigate the large deviation of the $k$ symbol empirical measure
and prove that the rate function can be obtained explicitly if $k\geq2$. Thus filling the gap between level 2 and 3.

However, in discrete time the extended process $(X_{t},X_{t+1},...,X_{t+k-1})$
is itself a Markov chain and therefore the intermediate level can be
derived from the Level 2. This magnification trick is no longer possible
in continuous time.
Until recently, no result existed in the literature to fill this Level 2-3 gap for continuous time.
The first study of this gap in the continuous time setting
was by Kesidis and Walrand \cite{Kesidis}, for pure jump processes
with two states. They obtained explicitly the rate function for the
joint probability of the empirical density and the empirical flow
counting the number of jumps between pair of states up to time T.
This intermediate level was then called 2.5. This issue was later
studied by De La Fortelle \cite{DeLaFortelle}, who obtained a weak
large deviation in the same context but for countable space.

Somewhat in parallel, in nonequilibrium statistical physics, it has been found  that
the empirical density at level 2 is not sufficient to study fluctuations
of the entropy production and of currents. This also motivated the search of an intermediate
level for pure jump and diffusion processes, by Maes and
collaborators \cite{maes0,Mae1}, and by Chernyak et al. \cite{Chertkov}.
Finally, Bertini et al. \cite{Bertini} succeeded in proving rigorously
the level 2.5 for pure jump processes in a countable space. Rigorous results for diffusion processes have been obtained in \cite{Kusuoka}.

The purpose of our contribution is to present the level 2.5 of large deviations for continuous time processes and discuss its connection with fluctuation relations. 
Whereas the explicit rate functions for the level 2.5 of large deviations calculated here have been obtained in 
\cite{DeLaFortelle,maes0,Chertkov,Mae1,Bertini}, our presentation unifies the proofs for pure jump and diffusion processes,
and clearly compares the two different methods used to obtain these rate functions, namely, tilting and a spectral method. 
Moreover, some of the proofs presented here are completely original.

The organization of the paper is as follows. Section \ref{sec:modobs} sets the stage with the definition of Markov processes, which include jump and diffusion processes. Particularly,
in subsection \ref{sub:Markov} we recall basic concepts of Markov processes like transition probability, generator, stationary and equilibrium  states, and trajectorial measure.
In subsection \ref{sub:eo}, we introduce the empirical density, empirical flow, empirical current, and the action functional, which are the fluctuating observables studied in the paper. 
In section \ref{sec:ftFR} we obtain the finite time fluctuation relation, which results as a tautology from the definition of the action functional. 
Section \ref{sec:2.5} is the cornerstone of the paper and deals with the level 2.5 of large deviations. In subsection \ref{sub:2,5Tilting} we use the tilting method to obtain the rate function characterizing the 
level 2.5. This proof is related to results from \cite{maes0,Mae1},  but the presentation given here is original. 
Subsection \ref{sub:2.5s} contains the spectral method. In this case, for pure jump processes our proof is original. For diffusion processes the spectral method has been used in \cite{Chertkov}, 
in comparison to this reference we expurgate the field theoretical language by using the Girsanov lemma. Finally, in section \ref{sec:ltFR} we obtain a stationary fluctuation relation at the level 2.5 and, by contraction,  
the GCEM symmetry for the fluctuating entropy.

\section{Models and Observables}

\label{sec:modobs}

\subsection{Homogeneous ergodic Markov processes}

\label{sub:Markov}

We start with a brief overview of homogeneous Markov processes
\cite{Chu1,Rev1,Rogers,Str}, considering continuous time Markov
processes $X_{t}$ taking values in a state space $\mathcal{E}$, which
can be continuous, as for example $\mathbb{R}^{d}$,  or a counting 
space.

\subsubsection{Elements of ergodic Markov processes}

\label{sub:eM} A time-homogeneous Markov process  can be defined by a family
of transitions kernel $P_{t}(x,dy)$, which is the conditional probability that $X_{t+t'}\in [y,y+dy]$ given that $X_{t'}= x$. This conditional probability satisfies the Chapmann-Kolmogorov rule 
\begin{equation}
\int_{\mathcal{E}}P_{s}(x,dy)P_{t}(y,dz)\,=\, P_{s+t}(x,dz),\label{CK}
\end{equation}
where the measure $dy$ means the Lebesgue measure or the counting measure, depending on $\mathcal{E}$.
The semi-group associated with the transition kernel is defined by its action on a bounded measurable function
$f$ in $\mathcal{E}$, 
\begin{equation}
P_{t}[f](x)\equiv\int_{\mathcal{E}} P_{t}(x,dy)f(y).
\end{equation}
The infinitesimal generator $L$, formally defined as $P_t\equiv \exp\left(tL\right)$, leads to the forward and
backward Kolmogorov equations,  
\begin{equation}
\partial_{t}P_{t}=P_{t}\circ L\qquad\textrm{and}\qquad\partial_{t}P_{t}=L\circ P_{t},\label{FK}
\end{equation}
respectively. The symbol $\circ$
means composition of operators and the initial condition is $P_{0}=\mathcal{I}$, where $\mathcal{I}$ is the identity kernel.  Conservative processes (without death or explosion), 
for which the normalization condition $\int P_{t}(x,dy)=1$ holds, are often considered in Physics. The generator must then obey $L[1]=0$, where $1$ is the function which is equal to $1$ on $\mathcal{E}$.

The time evolution of the instantaneous one point measure $\mu_{t}(dy)=\int_{\mathcal{E}}\mu_{0}(dx_{0})P_{t}(x_{0},dy)$
can be deduced from the Kolmogorov equation (\ref{FK}), leading to the Fokker-Planck equation $\partial_{t}\mu_{t}=L^{\dagger}[\mu_{t}]$,
where $L^{\dagger}$ is the adjoint of $L$ with respect to
the Lebesgue or counting measure. Since we are considering ergodic Markov processes, there is a unique invariant probability measure $\mu_{inv}$ satisfying 
\begin{equation}
L^{\dagger}[\mu_{inv}]=0.\label{INV-1}
\end{equation}
The process is said to be in \textbf{equilibrium} w.r.t $\mu_{inv}$ if the detailed balance relation is satisfied, i.e., 
\begin{equation}
\mu_{inv}(dx)P_{t}(x,dy)=\mu_{inv}(dy)P_t(y,dx).\label{DB-1}
\end{equation}
In the following it is assumed that the one point measure is smooth with respect to the Lebesgue measure, for example with the conditions of the Hormander theorem \cite{Hor,Mal}
for a diffusion process, leading to $\mu_{t}(dx)\equiv\rho_{t}(x)dx.$
With $\mu_{inv}(dx)\equiv\rho_{inv}(x)dx$, the detailed balance condition (\ref{DB-1})
can be written as 
\footnote{The expression $\rho_{inv}\circ L\circ\rho_{inv}^{-1}$ must be understood
as the composition of three operators, first the operator multiplication
by $\rho_{inv}^{-1},$ second the operator $L$ and last the operator
multiplication by $\rho_{inv}$. This type of notation is recurrently used
in the article.}
\begin{equation}
\rho_{inv}\circ L\circ\rho_{inv}^{-1}=L^{\dagger}.\label{DB'-1}
\end{equation}

In addition to the characterization by the semi-group or the generator,
a Markov process can be characterized by its trajectorial measure.
The sample path of the process up to time $T$ is the random function
$X_0^T$ : $t\in\left[0,T\right]\rightarrow X_{t}$ . It
is a random variable in the space of trajectories $D\left([0,T],\mathcal{E}\right)$.
This trajectorial measure $d\mathbb{P}_{L,\mu_{0},T}[x_0^T]$, where $\mu_0$ is the initial measure, 
is roughly the probability that the trajectory $X_0^T$ equals $x_0^T$.
The expectation of an arbitrary functional $F\left[X_0^T\right]$ of the trajectories is then written as,
\begin{equation}
\mathbb{E}_{L,\mu_{0}}\left[F\right]\ =\ \int_{\mathcal{}}F[x_0^T]\, d\mathbb{P}_{L,\mu_{0},T}[x_0^T].\label{eq:traj}
\end{equation}
 The finite time distributions are sufficient to characterize $d\mathbb{P}_{L,\mu_{0},T}$, more precisely, equation (\ref{eq:traj}) may be rewritten as
\begin{align}
\mathbb{E}_{L,\mu_{0}}\left[F\right]= &\int_{\mathcal{E}^{n+1}}F(x_{0},x_{1},...,x_{n-1},x_{n})\mu_{0}(dx_{0})\exp\left(t_{1}L\right)(x_{0},dx_{1})\label{eq:cyl}\\
&\times\exp\left(\left(t_{2}-t_{1}\right)L\right)(x_{1},dx_{2})....\exp\left(\left(T-t_{n-1}\right)L\right)(x_{n-1},dx_{n}),\nonumber
\end{align}
 for the cylindrical functional
\begin{equation}
F\left[X\right]=F(X_{0},X_{t_{1}},X_{t_{2}},...,X_{t_{n-1}},X_{T}), \label{eq:cy}
\end{equation}
with $0\leq t_{1}\leq t_{2}\leq....\leq t_{n-1}\leq T$. In the following we consider the two most prominent classes of Markov processes: jump and diffusion processes.

\subsubsection{Pure jump processes}

\label{sub:pjump}

A Markov process is called a pure jump process if after ``arriving''
into a state the system stays there for a random exponentially
distributed time interval and then jumps to another state.   The transition rates
$W(x,y)$  give the probability per unit of time for the transition
$x\rightarrow y$. Moreover, with regularity conditions (see chapter 8 of \cite{ethier} for example), it is possible to prove that for pure jump possesses the generator acting on
the bounded measurable function $h:\mathcal{E}\rightarrow\mathbb{R}$ is
\begin{equation}
L\left[h\right](x)=\int_{\mathcal{E}}W(x,y)\left(h(y)-h(x)\right)dy,\label{eq:genps}
\end{equation}
for all $x\in\mathcal{E}$. The detailed balance condition (\ref{DB-1}) with respect to the density $\rho_{inv}$ takes
the form
\begin{equation}
\rho_{inv}(x)W(x,y)=\rho_{inv}(y)W(y,x).\label{eq:BDS}
\end{equation}

A  relevant quantity in this paper is the current associated with the density $\rho_{t}$,
\begin{equation}
J_{\rho_{t}}(x,y)\equiv \rho_{t}(x)W(x,y)-W(y,x)\rho_{t}(y).\label{KDP}
\end{equation}
From equation (\ref{INV-1}),  the current associated
with the invariant density is conserved,  
\begin{equation}
\int dyJ_{\rho_{inv}}(x,y)=0.\label{eq:asc}
\end{equation}
At the trajectory level it is possible to compare the trajectorial measure
(\ref{eq:traj}) of two processes with different transition rates, with
the condition that they both have the same set of non vanishing rates. To this end,
we introduce the non conservative Markovian generator%
\footnote{In operational notation $L_{V_{1},V_{2}}=W\exp\left(V_{2}\right)-W\left[1\right]+V_{1}$.} 
\begin{equation}
L_{V_{1},V_{2}}\left[h\right](x)\equiv\left(\int_{\mathcal{E}}W(x,y)\left[\exp\left(V_{2}(x,y)\right)h(y)-h(x)\right]dy\right)+V_{1}(x)h(x),\label{eq:a1}
\end{equation}
for all functions $h$, $\textrm{with \ensuremath{V_{1}}: \ensuremath{\mathcal{E}\rightarrow\mathbb{R}} and \ensuremath{V_{2}:\mathcal{E}^{2}\rightarrow\mathbb{R}.}}$
We call this generator the twisted generator. From  the Girsanov
lemma \cite[Proposition 2.6]{KipnisLandim} and the Feynamn Kac relation
\cite{Rev1,Rogers}, it follows that $d\mathbb{P}_{L_{V_{1},V_{2}},\mu_{0},T}$
is absolutely continuous w.r.t. $d\mathbb{P}_{L,\mu_{0},T}$, and the
explicit Radon Nykodym derivative is given by 
\begin{equation}
\frac{d\mathbb{P}_{L_{V_{1},V_{2}},\mu_{0},T}}{d\mathbb{P}_{L,\mu_{0},T}}\left[x_0^T\right]=\exp\left(\sum_{0\leq s\leq T/ x_{s^-}\neq x_{s^+}}V_{2}(x_{s^{-}},x_{s^{+}})+\int_{0}^{T}dsV_{1}(x_{s})\right),\label{eq:a11}
\end{equation}
where $x_{s^-}\equiv\lim_{\delta\to0}x_{s-\delta}$ and $x_{s^+}\equiv\lim_{\delta\to0}x_{s+\delta}$. Hence, the sum $\sum_{0\leq s\leq T/ x_{s^-}\neq x_{s^+}}$ is over all jumps in the trajectory $x_0^T$.
In particular, for two conservative jump processes, one with rates $W$ and the other with rates $W_{V_{2}}(x,y)= W(x,y)\exp\left(V_{2}(x,y)\right)$ relation  (\ref{eq:a11}) becomes 
\begin{equation}
\frac{d\mathbb{P}_{L_{V_{2}},\mu_{0},T}}{d\mathbb{P}_{L,\mu_{0},T}}\left[x\right]=\exp\left(\sum_{0\leq s\leq T/ x_{s^-}\neq x_{s^+}}V_{2}(x_{s^{-}},x_{s^{+}})-\int_{0}^{T}ds\left(W\exp\left(V_{2}\right)-W\right)\left[1\right](x_{s})\right),\label{eq:a11-1}
\end{equation}
where $L_{V_{2}}$ is the conservative generator obtained from (\ref{eq:a1}) by setting 
\begin{equation}
V_{1}=\left(W\right)\left[1\right]-\left(W\exp\left(V_{2}\right)\right)\left[1\right]= \int W(x,y)dy-\int W(x,y)\exp(V_2(x,y))dy.
\end{equation}

\subsubsection{Diffusion processes}

\label{sub:pdiff}

A diffusion process $\, X_{t}\,$ in a $d$-dimensional manifold is described
by the differential equation
\begin{equation}
dX=A_{0}(X)dt+\sum_\alpha A_{\alpha}(X)\circ dW_{\alpha}(t).\label{eq:diff}
\end{equation}
where the drift $A_{0}$ and the diffusion coefficient $A_{\alpha}$ are arbitrary smooth vector fields
on $\mathcal{E}$, $W_{\alpha}$ are independent Wiener processes, and the range of $\alpha$ is model dependent.
The symbol $\circ$ indicates that the Stratonovich convention
is used. The explicit form of the generator related to (\ref{eq:diff}) is 
\begin{equation}
L=A_{0}.\nabla+\sum_\alpha\frac{1}{2}\left(A_{\alpha}.\nabla\right)^{2}=\widehat{A_{0}}.\nabla+\frac{1}{2}\nabla.D.\nabla,\label{eq:gpd}
\end{equation}
with the modified drift and covariance
\begin{equation}
\widehat{A_{0}}(x)=A_{0}(x)-\frac{1}{2}\sum_\alpha\left(\nabla.A_{\alpha}\right)(x)A_{\alpha}(x)\qquad\textrm{and}\qquad D^{ij}(x)=\sum_\alpha A_{\alpha}^{i}(x)A_{\alpha}^{j}(x),\label{eq:diffel}
\end{equation}
respectively, where $i=1,\ldots,d$ and $j=1,\ldots,d$. It is assumed that $D$ is strictly positive. The detailed balance relation (\ref{DB-1}) with respect to the invariant measure $\mu_{inv}(dx)=\rho_{inv}(x)dx$
is then equivalent to $\widehat{A_{0}}=\frac{D}{2}\nabla\left(\ln\rho_{inv}\right)$.
 
A central quantity for diffusion processes is the hydrodynamic
probability current \cite{risken}
\begin{equation}
J_{\rho_{t}}=\widehat{A}_{0}\rho_{t}-\frac{D}{2}(\nabla\rho_{t})\label{jv}.
\end{equation}
The conservation of the current associated with the invariant density then reads
\begin{equation}
\nabla.J_{\rho_{inv}}=0.\label{eq:tcdp}
\end{equation}

Similar to jump processes, the trajectorial measure
of two diffusion processes can be compared with a generator corresponding to a non-conservative process, which in the present case is defined as
\begin{equation}
L'\equiv L+B_2.\nabla+B_1,\label{eq:a2},
\end{equation}
where $B_2$ and $B_1$ are arbitrary vector field and  scalar, respectively.
Combining the Cameron-Martin-Girsanov
lemma \cite{Rogers,Str} and  the Feynamm-Kac
relation \cite{Rev1,Rogers}, it follows that 
\begin{equation}
\frac{d\mathbb{P}_{L',\mu_{0},T}\left[x\right]}{d\mathbb{P}_{L,\mu_{0},T}\left[x\right]}=\exp(V_{T}\left[x\right]),\label{eq:a22}
\end{equation}
where 
\begin{align}
V_{T}=\int_{0}^{T}\left[D^{-1}(x_{u})B_2(x_{u})\circ dx_{u}
 +\left(B_1(x_{u})-D^{-1}(x_{u})B_2(x_{u})\left(\widehat{A_{0}}+\frac{B_2}{2}\right)(x_{u})-\frac{1}{2}\left(\nabla.B_2\right)(x_{u})\right)du\right].\nonumber\\
\end{align}
Choosing 
\begin{equation}
B_2=DV_{2}\qquad\textrm{and}\qquad B_1=V_{2}.\left(\widehat{A_{0}}+\frac{DV_{2}}{2}\right)+\frac{1}{2}\nabla.\left(DV_{2}\right)+V_{1},
\end{equation}
we obtain 
\begin{equation}
V_{T}=\int_{0}^{T}dt\left[V_{1}(X_{t})+V_{2}(X_{t})\circ dX_{t}\right].
\end{equation}
Equation (\ref{eq:a22}) then becomes
\begin{equation}
\frac{d\mathbb{P}_{L_{V_{1},V_{2}},\mu_{0},T}}{d\mathbb{P}_{L,\mu_{0},T}}\left[X\right]=\exp\left(\int_{0}^{T}dt\left[V_{1}(X_{t})+V_{2}(X_{t})\circ dX_{t}\right]\right),\label{eq:a22222}
\end{equation}
where the twisted generator reads
 \begin{align}
L_{V_{1},V_{2}} &= L'=  L+DV_{2}\nabla+V_{2}.\left(\widehat{A_{0}}+\frac{DV_{2}}{2}\right)+\frac{1}{2}\nabla.\left(DV_{2}\right)+V_{1}\nonumber\\
&=\widehat{A_{0}}.\left(\nabla+V_{2}\right)+\left(\nabla+V_{2}\right)\circ\frac{D}{2}\circ\left(\nabla+V_{2}\right)+V_{1}.
\label{eq28}
\end{align}

\subsection{Empirical observables and ergodic behavior}

\label{sub:eo}

\subsubsection{Empirical density, flow and current}
\label{sub:ed}

The set of functional observables that define the Level 2.5 of large deviations depend of the type of Markov processes considered.
For pure jump processes the set of observables is the empirical density $\rho_{T}^{e}$ and empirical flow $C_{T}^{e}$. They are given by
 \begin{equation}
\rho_{T}^{e}(x)=\frac{1}{T}\int_{0}^{T}\delta\left(X_{t}-x\right)dt\qquad\textrm{and}\qquad C_{T}^{e}(x,y)=\frac{1}{T}\sum_{0\leq s\leq T/ X_{s^-}\neq X_{s^+}}\delta\left(X_{t^{-}}-x\right)\delta\left(X_{t^{+}}-y\right).
\label{eq:eops}
\end{equation}
The empirical density $\rho_{T}^{e}(x)$ 
\footnote{Rigorously, we should instead define the empirical measure $\mu_{T}^{e}=\frac{1}{T}\int_{0}^{T}\delta_{X_{t}}dt$.} can be understood as the fraction of time spent in $x$ over $\left[0,T\right]$ and the empirical flow $C_{T}^{e}(x,y)$  as
the number of jumps from $x$ to $y$ (times $1/T$) during the trajectory. Another functional of central interest is the empirical current  
\begin{equation}
J_{T}^{e}(x,y)=C_{T}^{e}(x,y)-C_{T}^{e}(y,x).\label{eq:ecjp}
\end{equation}
Since we assume the system to be ergodic,  the law of large numbers for the empirical density and flow becomes  
\begin{equation}
\rho_{T}^{e}\rightarrow\rho_{inv}\qquad\textrm{and}\qquad C_{T}^{e}\rightarrow C_{\rho_{inv}},\label{eq:teops}
\end{equation}
where 
\begin{equation}
C_{\rho_{inv}}(x,y)=\rho_{inv}(x)W(x,y).\label{eq:def}
\end{equation}
Moreover, the finite time Kirchkoff's law \cite{Kirchkoff} reads
\begin{align}
& \int dyC_{T}^{e}(x,y)-\int dyC_{T}^{e}(y,x)  =  \frac{1}{T}\sum_{0\leq s\leq T/ X_{s^-}\neq X_{s^+}}\delta\left(X_{t^{-}}-x\right)-\frac{1}{T}\sum_{0\leq s\leq T/ X_{s^-}\neq X_{s^+}}\delta\left(X_{t^{+}}-x\right)\nonumber\\
 & =  \frac{\delta\left(X_{0}-x\right)-\delta\left(X_{T}-x\right)}{T}=\textrm{O}(1/T).\label{eq:cpps}
\end{align}
In the following we will show that the large deviation rate function of  $C_{T}^{e}$ is infinite  for any untypical $C$ not fulfilling
\begin{equation}
\int dyC(x,y)=\int dyC(y,x).\label{eq:cpj}\
\end{equation}

For diffusion processes, the set of observables is composed by the
empirical density $\rho_{T}^{e}$ and the empirical current $j_{T}^{e}$, which read
\begin{equation}
\rho_{T}^{e}(x)=\frac{1}{T}\int_{0}^{T}\delta\left(X_{t}-x\right)dt\text{ }\qquad\textrm{and}\qquad j_{T}^{e}(x)=\frac{1}{T}\int_{0\text{ }}^{T}\delta\left(X_{t}-x\right)\circ dX_{t}.\label{eq:eopd}
\end{equation}
Roughly speaking, the empirical current (see \cite{Flandoli} for a rigorous definition) is the sum of the displacements
that the system makes if it is at $x$. For diffusion processes, with the ergodic assumption the law of large numbers takes the form
\begin{equation}
\rho_{T}^{e}\rightarrow\rho_{inv}\qquad\textrm{and}\qquad j_{T}^{e}\rightarrow J_{\rho_{inv}}.\label{eq:teopd}
\end{equation}
where the current $J_{\rho_{inv}}$ is defined in relation (\ref{jv}).
From the definition (\ref{eq:eopd}), we obtain the pathwise constraint
\footnote{
\begin{eqnarray*}
\int_{\mathcal{E}}dxg(x)\nabla.j_{T}^{e}(x) & = & -\int_{\mathcal{E}}dxj_{T}^{e}(x).\nabla g(x)=-\frac{1}{T}\int_{0\text{ }}^{T}\nabla g(X_{t})\circ dX_{t}=\frac{1}{T}\left(g(X_{0})-g(Xt)\right),\qquad\textrm{for all functions $g$}.
\end{eqnarray*}}
\begin{equation}
\nabla.j_{T}^{e}(x)=\frac{1}{T}\left(\delta\left(X_{0}-x\right)-\delta\left(X_{t}-x\right)\right).\label{eq:cppd}
\end{equation}
Hence, analogously  to (\ref{eq:cpj}) the large deviation rate function of $j_{T}^{e}$ is infinite at any $j$ not fulfilling 
\begin{equation}
\nabla.j=0.\label{eq:cpd}
\end{equation}

\subsubsection{Action functional and fluctuating entropy}
\label{sub:af}

For time-homogeneous processes, the action functional $\mathbb{W}_{T}$ is obtained by comparing the trajectorial
measure of $X_{t}$ with the time-reversed trajectorial measure.
At the level of trajectories, we introduce the path-wise time inversion
\footnote{Here, we do not consider the case where the time inversion acts non-trivially
on the space $\mathcal{E}$. For example, such a situation takes place
for the non-over-damped Kramers equation \cite{Che1}.}
$R$ acting  on the space of trajectories as 
\begin{equation}
R\left[X_0^T\right]_{t}\equiv\left[X_0^T\right]_{T-t},\label{eq:Rtraj}
\end{equation}
where $\left[X_0^T\right]_{t}\equiv X_t$.

The action functional is defined by the relation
\begin{equation}
\exp\left(-\mathbb{W}_{T}\right)\equiv\frac{R_{\star}\left(d\mathbb{P}_{L,\mu_{0}^{b},T}\right)}{d\mathbb{P}_{L,\mu_{0},T}}.
\label{actionfunct}
\end{equation}
where $\mu_{0}^{b}$ is the arbitrary initial measure of the reversed trajectory and 
the push-forward measure can be loosely written as $R_{\star}\left(d\mathbb{P}_{L,\mu_{0}^{b},T}\right)[x_0^T]= d\mathbb{P}_{L,\mu_{0}^{b},T}\left[R\left[x_0^T\right]\right]$.
Due to the freedom in choosing $\mu_{0}$ and  $\mu_{0}^{b}$, it is possible to identify the action
functional $\mathbb{W}_{T}$$ $ with different quantities. It becomes
the fluctuating total entropy production $\mathbb{\sigma}_{T}$ for 
$\mu_{0}^{b}(dx)=\mu_{T}(dx)\equiv\int dy\rho_{0}(y)P_{0}^{T}(y,x)dx$ and the fluctuating entropy increase of the external environment $\mathbb{J}_{T}$ for
$\mu_{0}(dx)=\mu_{0}^{b}(dx)=dx$. The difference between $\mathbb{\sigma}_{T}$ and  $\mathbb{J}_{T}$ is the boundary term $\ln\left(\rho_{0}(x_{0})\right)-\ln\left(\rho_{T}(x_{T})\right)$, which is the variation of the entropy of the system.
We note that names like total entropy production or entropy increase of the external environment become meaningful only if a Markov process is given a clear 
physical interpretation. In this case these functionals are related to key thermodynamic quantities \cite{Sei}.

For pure jump processes this action functional is \cite{Maes99,LebowSp}
\begin{equation}
\mathbb{W}_{T}=\ln\left(\rho_{0}(X_{0})\right)-\ln\left(\rho_{0}^{b}(X_{T})\right)+\sum_{0\leq s\leq T/ X_{s^-}\neq X_{s^+}}\ln\left[\frac{W(X_{t^{-}},X_{t^{+}})}{W(X_{t^{+}},X_{t^{-}})}\right].\label{eq:Wjp}
\end{equation}
For  diffusion processes it reads \cite{LebowSp}
\begin{equation}
\mathbb{W}_{T}=\ln\left(\rho_{0}(X_{0})\right)-\ln\left(\rho_{0}^{b}(X_{T})\right)+2\int_{0}^{T}dt\widehat{A_{0}}\left(X_{t}\right).D^{-1}\left(X_{t}\right)\circ dX_{t}.\label{eq:Wdp}
\end{equation}

\section{Transient Fluctuation Relation}
\label{sec:ftFR}

Before obtaining the rate function at the level 2.5, let us briefly discuss the transient fluctuation relation.
From the definition of the action functional (\ref{actionfunct}) it follows that for all functionals $F_{\left[0,T\right]}$, 
\begin{equation}
\mathbb{E}_{\mu_{0}^{b},L}\left[F_{\left[0,T\right]}\circ R\right]=\mathbb{E}_{\mu_{0},L}\left[F_{\left[0,T\right]}\exp\left(-\mathbb{W}_{T}\right)\right].\label{fluC}
\end{equation}
The backward action functional is defined as
 \begin{equation}
\exp\left(-\mathbb{W}_{T}^{b}\right) \equiv \frac{R_{\star}\left(d\mathbb{P}_{L,\mu_{0},T}\right)}{d\mathbb{P}_{L,\mu_{0}^{b},T}}.\label{actionfunct-3}
\end{equation}
Comparing (\ref{actionfunct}) and (\ref{actionfunct-3}) we obtain the antisymmetric relation
 \begin{equation}
\mathbb{W}_{T}^{b}=-\mathbb{W}_{T}\circ R.\label{eq:antW}
\end{equation}
For the special case $F_{\left[0,T\right]}=\delta(\mathbb{W}_{T}-W)$, with $\delta$ denoting the indicator function, relation (\ref{fluC}) becomes the generalized Crooks relation \cite{Crooks2,Maes99,LebowSp,Che1,Sei}
\begin{equation}
\mathbb{P}_{\mu_{0}^{b},L}(\mathbb{W}_{T}^{b}=-W)=\exp\left(-W\right)\mathbb{P}_{\mu_{0},L}(\mathbb{W}_{T}=W).
\label{eq:crooks}
\end{equation}
From (\ref{fluC}), we also deduce the Jarzynski equality \cite{Crooks2,Jarz} 
\begin{equation}
\mathbb{E}_{\mu_{0},L}\left[\exp(-\mathbb{W}_{T})\right]=1.\label{eq:Jar}
\end{equation}
This relation implies two important results. First, (\ref{eq:Jar}) and Jensen's inequality
gives the second law of thermodynamics $\mathbb{E}_{\mu_{0},L}\left[\mathbb{W}_{T}\right]\geq0$.
Second, (\ref{eq:Jar}) and the Markov inequality $\mathbb{P}_{\mu_{0},L}\left(\exp\left(-\mathbb{W}_{T}\right)\geq\exp(L)\right)\leq\frac{\mathbb{E}_{\mu_{0},L}\left[\exp(-\mathbb{W}_{T})\right]}{\exp(L)}$ gives an upper bound
\footnote{A better upper bound has been obtained in \cite{Chetrite} using the classical
Martingale inequality.}
on the probability of ``transient deviations'' from the second law, i.e., $\mathbb{P}_{\mu_{0},L}\left(\mathbb{W}_{T}\leq-L\right)\leq\exp\left(-L\right).$

\section{Heuristic proof for 2.5 large deviations}
\label{sec:2.5}

In this section we demonstrate that
the joint fluctuation of empirical density and empirical flow for jump 
processes, and the joint fluctuation of empirical density
and empirical current for diffusion processes admit a large
deviation regime with an explicit rate function.
For jump processes this rate function reads \cite{maes0}
\begin{equation}
 I\left[\rho,C\right] =\begin{cases}
\int dxdy\left(\begin{array}{c}
-C(x,y)+\rho(x)W(x,y)\\
+C(x,y)\ln\frac{C(x,y)}{\rho(x)W(x,y)}\end{array}\right) \qquad\text{if }\forall x\in\mathcal{E}:\int dyC(x,y)=\int dyC(y,x)\\
\infty \qquad \textrm{otherwise,}\end{cases}\label{eq:2,5jp}
\end{equation}
while for diffusion processes it is \cite{Mae1,Chertkov}
\begin{equation}
I\left[\rho,j\right]=\left\{ \begin{array}{c}
\frac{1}{2}\int dx(j-J_{\rho})(\rho D)^{-1}(j-J_{\rho})\qquad\text{if }\nabla.j=0\\
\infty\qquad\text{otherwise.}\end{array}\right.\label{eq:2,5dp}
\end{equation}
Note that the constraints $\int dyC(x,y)=\int dyC(y,x)$ and $\nabla.j=0$ come from (\ref{eq:cpps}) and (\ref{eq:cppd}), respectively.
Formally, by contraction we can obtain the Donsker-Varadhan
variational expression for the rate function for the level 2 of large deviations from the level 2.5 rate function. Explicitly, for pure jump processes  $I(\rho)=\min_{C}\left[I(\rho,C)\right]$,
whereas  for diffusion processes $I(\rho)=\min_{j}\left[I(\rho,j)\right]$. These relations lead to
\begin{equation}
I\left[\rho\right]=-\inf_{\left[h\right]>0}\left[\int dx\rho(x)h^{-1}(x)L\left[h\right](x)\right],\label{eq:LD2}
\end{equation}
where the minimization is over strictly positive functions $h$. A rigorous
proof of this contraction for pure jump processes can be found in \cite{Ber2}.
Similarly, a formal contraction implies that the action functional (\ref{eq:Wjp}) (or (\ref{eq:Wdp}) for diffusion processes)
fulfills a Large Deviation principle. It is also possible
to obtain the rate function related to the joint probability of the empirical density $\rho_{T}^{e}(x,y)$ and the empirical current $J_{T}^{e}(x,y)$ by contraction from (\ref{eq:2,5jp}) \cite{maes0}.

We present two methods to prove (\ref{eq:2,5jp}) and (\ref{eq:2,5dp}): tilting and a spectral method. The proof for jump processes using the spectral method is original. Proofs using tilting 
for pure jump processes can be found in \cite{maes0} and for diffusion processes in \cite{Mae1}. Another proof for diffusion  processes using the spectral method was obtained in \cite{Chertkov}. 
The novelty in these cases is in our presentation, which highlight the generality of both methods. A third method, which is totally rigorous, for pure jump processes in a countable space related 
to the contraction of the rate function of the level 3 of large deviations has been recently obtained  in \cite{Bertini}. Whereas the proof using the tilting method is more direct, in the spectral
method a connection between the rate function and the maximum eigenvalue of a modified generator is established. This connection is often useful for numerical calculations of rate functions.

\subsection{Tilting}
\label{sub:2,5Tilting}

We consider, for  general
stochastic processes $X_{t}$, \footnote{ $X_{t}$ does not need to be Markovian here.}
the joint large deviation
of $N$ observables $\mathit{\overrightarrow{\omega_{t}^{e}}}\equiv$$\left\{ \omega_{t,1}^{e},\omega_{t,2}^{e},.....,\omega_{t,N}^{e}\right\} $.
The trajectorial measure is denoted by $d\mathbb{P}_{\mu_{0},T}$ and $\overrightarrow{\omega_{inv}}\equiv\left\{\omega_{inv,1},\omega_{inv,2},.....,\omega_{inv,N}\right\} $
represents the typical behavior of $\mathit{\overrightarrow{\omega_{t}^{e}}}$, with typical behavior meaning almost sure convergence. If the following two conditions are satisfied  then the family of 
probability measures $\left(\mathbb{P}_{\mu_{0},T}\circ\left\{ \mathit{\overrightarrow{\omega_{t}^{e}}}\right\} ^{-1}\right)_{t\geq0}$, or equivalently $\mathit{\overrightarrow{\omega_{t}^{e}}}$, satisfies 
a large deviation principle with rate function
$I\left(\mathit{\overrightarrow{\omega}}\right)$, where $\mathit{\overrightarrow{\omega}}=\left\{\omega_{1},\omega_{2},.....,\omega_{N}\right\} $ is the desired untypical behavior.
\begin{itemize}
\item \textbf{Condition 1}: There exists an ergodic tilted process $X_{t}'$, with trajectorial measure
$d\mathbb{P}'_{\mu_{0},T}$, such that its typical behavior is
$\mathit{\overrightarrow{\omega_{t}^{e}}}$.
\item \textbf{Condition 2}: For this tilted process, there exists a function $I$ defined by the asymptotic relation 
$\frac{d\mathbb{P}_{\mu_{0},T}}{d\mathbb{P}'_{\mu_{0},T}}\left[X\right]\sim\exp\left(-TI\left(\mathit{\overrightarrow{\omega_{T}^{e}}}\right)\right)$.
This means that asymptotically the Radon-Nykodym derivative can be expressed in terms of the $N$ observables $\omega_{t,1}^{e},\omega_{t,2}^{e},.....,\omega_{t,N}^{e}$.
\end{itemize}
Note that larger $N$ makes the fulfillment of the first condition harder, while the fulfillment of second condition becomes easier. For a fixed process
$X_{t}$ and a fixed observable $\mathit{\overrightarrow{\omega_{t}^{e}}}$, we postulate that the process $X_t'$ exists.

\paragraph{Formal proof : }

From the second condition it follows that
\begin{align}
& \mathbb{P}_{\mu_{0},T}\left[\mathit{\overrightarrow{\omega_{T}^{e}}}\simeq\overrightarrow{\omega}\right] =  \int d\mathbb{P}_{\mu_{0},T}\left[X\right]\delta(\mathit{\overrightarrow{\omega_{T}^{e}}}-\overrightarrow{\omega})=\int d\mathbb{P}'_{\mu_{0},T}\left[X\right].\frac{d\mathbb{P}_{\mu_{0},T}}{d\mathbb{P}'_{\mu_{0},T}}\left[X\right]\delta(\mathit{\overrightarrow{\omega_{T}^{e}}}-\overrightarrow{\omega})\nonumber\\
& \sim  \int d\mathbb{P}'_{\mu_{0},T}\left[X\right].\exp\left(-TI\left(\mathit{\overrightarrow{\omega_{T}^{e}}}\right)\right)\delta(\mathit{\overrightarrow{\omega_{T}^{e}}}-\overrightarrow{\omega})\sim\exp\left(-TI\left(\mathit{\overrightarrow{\omega}}\right)\right)\int d\mathbb{P}'_{\mu_{0},T}\left[X\right]\delta(\mathit{\overrightarrow{\omega_{T}^{e}}}-\overrightarrow{\omega}).\nonumber\\ 
\label{eq54}
\end{align}
Since the process  $X'_{t}$ is assumed to be ergodic, with the first condition, we obtain 
\begin{equation}
\int d\mathbb{P}'_{\mu_{0},T}\left[X\right]\delta(\mathit{\overrightarrow{\omega_{T}^{e}}}-\overrightarrow{\omega})=\mathbb{P}'_{\mu_{0}}\left[\mathit{\overrightarrow{\omega_{T}^{e}}}\simeq\overrightarrow{\omega}\right]\rightarrow1,\label{eq:tilth2}
\end{equation}
which, with (\ref{eq54}), gives the required Large deviation rate function $I$.
Rigorously, following the same procedure for $\mathbb{P}_{\mu_{0}}\left[\mathit{\overrightarrow{\omega_{T}^{e}}}\in B\left(\overrightarrow{\omega},\epsilon\right)\right]$,
where $B\left(\overrightarrow{\omega},\epsilon\right)$ an open ball of radius $\epsilon$,  the lower bound of the rate function (\ref{eq:Ldev}) is obtained \cite{Bertini}. 
We note that these two conditions are not enough for a rigorous proof, which requires a lower and an upper bound on the rate function \cite{Bertini,Kusuoka}. 

\paragraph{Examples : }
\begin{itemize}
\item If $X_{t}$ is a Markov process and $\mathit{\overrightarrow{\omega_{t}^{e}}}\equiv$$\left\{ \rho_{t}^{e}\right\}$,
from the Girsanov relation (\ref{eq:a11-1}) (or (\ref{eq:a22}) for diffusion processes), we obtain that it is not possible to find a process fulfilling the second condition. 
The solution to find an explicit rate function is then to increase $N$.
\item If $X_{t}$ is a pure jump process and $\mathit{\overrightarrow{\omega_{t}^{e}}}=\left\{ \rho_{t}^{e},C_{t}^{e}\right\}$,
by choosing $X'$ with the transition rates
\begin{equation}
W'(x,y)=\frac{C(x,y)}{\rho(x)},\label{eq:Wtilpj}
\end{equation}
the ergodic behavior of $X'_t$ becomes $\rho'_{inv}=\rho$ and $C_{\rho'_{inv}}=C$, which implies the fulfillment of condition 1.
The process $X'_t$ also obeys the conservation law (\ref{eq:asc}), leading to the constraint on the marginal of $C$ in the rate function (\ref{eq:2,5jp}).
The Girsanov relation (\ref{eq:a11-1}) with $V_{2}(x,y)=\ln\left(\frac{C(x,y)}{\rho(x)W(x,y)}\right)$ becomes
\begin{align}
& \frac{d\mathbb{P}_{L_{V_2},T}}{d\mathbb{P}_{L,T}}\left[X\right]\\
& =  \exp\left[\sum_{0\leq s\leq T/ X_{s^-}\neq X_{s^+}}\ln\left(\frac{C(X_{s^{-}},X_{s^{+}})}{\rho(X_{s^{-}})W(X_{s^{-}},X_{s^{+}})}\right)-\int_{0}^{T}ds\int_{\mathcal{E}}dy\left(\frac{C(X_{s},y)}{\rho(X_{s})}-W(X_{s},y)\right)\right]\label{eq:a11-1-1}\nonumber\\
 & = \exp\left[T\int_{\mathcal{E}^{2}}dydx\left[C_{T}^{e}(x,y)\ln\left(\frac{C(x,y)}{\rho(x)W(x,y)}\right)-\rho_{T}^{e}(x)\left(\frac{C(x,y)}{\rho(x)}-W(x,y)\right)\right]\right]\nonumber.
\end{align}
Hence, condition 2 is exactly verified at finite time with the rate function $I$ given by (\ref{eq:2,5jp}). 
\end{itemize}
\begin{itemize}
\item If $X_{t}$ is a diffusion process and $\mathit{\overrightarrow{\omega_{t}^{e}}}=\left\{ \rho_{t}^{e},j_{t}^{e}\right\}$,
condition 1 is fulfilled by choosing $X'_{t}$ with drift and diffusion coefficient 
\begin{equation}
A_{0}'=\frac{j+\frac{D}{2}\nabla\rho}{\rho}\qquad\textrm{and}\qquad A_{\alpha}'=A_{\alpha}.\label{eq:tdpe-2}
\end{equation}
This can be shown with the ergodic law (\ref{eq:eopd}), which implies
\begin{equation}
\rho'_{inv}=\rho\textrm{ and }J_{\rho'_{inv}}=j,\label{eq:ergdp}
\end{equation}
\end{itemize}
where $\rho'{}_{inv}$ is the invariant density of the process $X'_{t}$.
From the Girsanov relation (\ref{eq:a22}), condition 2 is verified with $I$ given by (\ref{eq:2,5dp}). 
\begin{itemize}
\item It is possible to apply the tilting method to find the rate function of more informative
quantities, e.g., the $m$-words generalization of empirical flow associated
with a pure jump process \cite{Che2}. The method can also be used to 
obtain the rate function of the empirical density and flow of pure jump processes that are non-homogeneous and periodic in time \cite{Ber3}.
\end{itemize}

\subsection{Spectral method}
\label{sub:2.5s}

\subsubsection{Generating function}

\label{sub:gf}

The scaled cumulant generating function associated with the vector $\mathit{\overrightarrow{\omega_{t}^{e}}}$
is defined as
\begin{equation}
\Lambda\left[V_{1},V_{2},...,V_{N}\right]=\lim_{T\rightarrow\infty}\frac{1}{T}\ln\left(\mathbb{E}_{\mu_{0},L}\left[\exp\left(T\sum_{i=1}^{N}\left\langle \omega_{t,i}^{e},V_{i}\right\rangle \right)\right]\right)\label{eq:fgcg}
\end{equation}
where $V_{i}$ are objects having the same tensorial nature as $\omega_{t,i}^{e}$ and $\left\langle .,.\right\rangle$ denotes the
associated canonical scalar product. Assuming that the G\"artner-Ellis theorem \cite{DenH,Dembo} is still valid in this functional form
\footnote{For a theoretical Physicist point of view, this theorem is a functional Laplace
transform followed by a saddle point approximation.},
then if $\Lambda$ exist and is differentiable for all $V_{i}$,  the family of probability measures $\left(\mathbb{P}_{\mu_{0},T}\circ\left\{ \overrightarrow{\omega_{t}^{e}}\right\} ^{-1}\right)_{t\geq0}$ satisfies a large deviation principle
with rate function 
\begin{equation}
I\left[\omega_{1},\omega_{2},...,\omega_{N}\right]=\sup_{\overrightarrow{V}}\left\{\sum_{i=1}^{N}\left\langle \omega_{i},V_{i}\right\rangle -\Lambda\left[V_{1},V_{2},...,V_{N}\right]\right\}.\label{eq:GE}
\end{equation}
For pure jump processes, with  $\mathit{\overrightarrow{\omega_{t}^{e}}}= \left\{ \rho_{t}^{e},C_{t}^{e}\right\}$, the scaled cumulant generating function becomes 
\begin{equation}
\Lambda\left[V_{1},V_{2}\right]=\lim_{T\rightarrow\infty}\frac{1}{T}\ln\left(\mathbb{E}_{\mu_{0},L}\left[\exp\left(\int_{0}^{T}dtV_{1}(X_{t})+\sum_{0\leq s\leq T/ X_{s^-}\neq X_{s^+}}V_{2}\left(X_{t^{-}},X_{t^{+}}\right)\right)\right]\right).\label{eq:fgcps}
\end{equation}
For diffusion processes, with $\mathit{\overrightarrow{\omega_{t}^{e}}}=\left\{ \rho_{t}^{e},j_{t}^{e}\right\}$ we obtain
\begin{equation}
\Lambda\left[V_{1},V_{2}\right]=\lim_{T\rightarrow\infty}\frac{1}{T}\ln\left(\mathbb{E}_{\mu_{0},L}\left[\exp\left(\int_{0}^{T}dt\left[V_{1}(X_{t})+V_{2}(X_{t})\circ dX_{t}\right]\right)\right]\right).\label{eq:fgcpd}
\end{equation}

\subsubsection{Twisted process }

\label{par:tp}
Defining 
\begin{equation}
A_{T}^{e}\equiv\frac{1}{T}\left(\int_{0}^{T}dtV_{1}(X_{t})+\sum_{0\leq s\leq T/ X_{s^-}\neq X_{s^+}}V_{2}\left(X_{t^{-}},X_{t^{+}}\right)\right), 
\end{equation}
relation (\ref{eq:a11}), which is valid for pure jump processes, is equivalent to
\begin{equation}
\mathbb{E}_{L,\mu_{0}}\left[\exp\left(TA_{T}^{e}\right)F\right]=\mathbb{E}_{L_{V_{1},V_{2}},\mu_{0}}\left[F\right],\label{eq:Fey}
\end{equation}
where $F$ is a generic functional and $L_{V_{1},V_{2}}$ is defined in (\ref{eq:a1}) for pure jump processes.  For diffusion processes  
\begin{equation}
A_{T}^{e}\equiv\frac{1}{T}\left(\int_{0}^{T}dt\left[V_{1}(X_{t})+V_{2}(X_{t})\circ dX_{t}\right]\right),
\end{equation}
and relation (\ref{eq:a22222}) is equivalent to (\ref{eq:Fey}), with $L_{V_{1},V_{2}}$ defined in (\ref{eq28}).

The special functional $F=\delta(X_{T}-y)$ gives the Feynamn-Kac type relation 
\begin{eqnarray}
\mathbb{E}_{L,\mu_{0}}\left[\exp\left(TA_{T}^{e}\right)\delta(X_{T}-y)\right] & = & \int_{\mathcal{E}}\mu_{0}(dx_{0})\exp\left(TL_{V_{1},V_{2}}\right)(x_{0},y).\label{eq:dvl-2}
\end{eqnarray}
We assume that the twisted operator $L_{V_{1},V_{2}}$ is of Perron-Frobenius type, i.e., there  exists a positive gaped principal eigenvalue
with maximal real part $\lambda\left[V_{1},V_{2}\right]$ related to a unique positive right eigenvector $r\left[V_{1},V_{2}\right]$ and a unique
positive left eigenvector $l\left[V_{1},V_{2}\right]$
\footnote{These properties follow from the Krein-Rutman theorem \cite{Krein}, which, however,
requires that the operator $L_{V_{1},V_{2}}$ is compact. For a uniformly elliptic operator
in divergent form as the generator of a diffusion process, a version
of the Krein-Rutman theorem is proven, for example, in chapter 6.5.2 of \cite{Evans}, 
where the hypothesis are: $\mathcal{E}$ is bounded, open and connected; $\partial\mathcal{E}$ is smooth;
$D$ and $\widehat{A_{0}}$ are smooth; $L_{V_{1},V_{2}}\left[1\right]\geq0$ on $\mathcal{E}$. 
Strictly speaking, the theorem is not valid if, for example, $\mathcal{E}$ is not bounded, with
the extension for an unbounded $\mathcal{E}$ being a difficult and contemporary
problem \cite{Ber}. Even though  we are not aware of proof for unbounded $\mathcal{E}$ in the mathematics literature, more sophisticated related results 
do exist, as for example in chapter 4.11 of \cite{Pin}.
From a physicist perspective, if the drift of the process is sufficiently confining
then the result for bounded $\mathcal{E}$ case should also be true for unbounded $\mathcal{E}$.
}.
Multiplicative factors are fixed by normalization as
\begin{equation}
\int_{\mathcal{E}}l\left[V_{1},V_{2}\right](x)dx=1\qquad\textrm{and}\qquad\int_{\mathcal{E}}l\left[V_{1},V_{2}\right](x)r\left[V_{1},V_{2}\right](x)dx=1.\label{eq:norm}
\end{equation}
It is also assumed that the initial measure fulfills
\begin{equation}
\int_{\mathcal{E}}\mu_{0}(dx)r\left[V_{1},V_{2}\right](x)<\infty.
\end{equation}
With this principal eigenvalue and its associated eigenvectors, the semi-group generated by $L_{V_{1},V_{2}}$ can be expanded as
\begin{equation}
\exp\left(TL_{V_{1},V_{2}}\right)(x,y)=\exp\left(T\lambda_{V_{1},V_{2}}\right)\left(r\left[V_{1},V_{2}\right](x)l\left[V_{1},V_{2}\right](y)+O\left(\exp\left(-t\Delta_{V_{1},V_{2}}\right)\right)\right),\label{eq:asysg}
\end{equation}
where $\Delta_{V_{1},V_{2}}$ is the spectral gap. Combining this last equation with the Feynman-Kac relation (\ref{eq:dvl-2}) we obtain 
\begin{equation}
\mathbb{E}_{\mu_{0},L}\left[\exp\left(TA_{T}^{e}\right)\delta(X_{T}-y)\right]=\exp\left(T\lambda_{V_{1},V_{2}}\right)\int_{\mathcal{E}}\mu_{0}(dx_{0})\left(r\left[V_{1},V_{2}\right](x)l\left[V_{1},V_{2}\right](y)+O\left(\exp\left(-t\Delta_{V_{1},V_{2}}\right)\right)\right).
\label{eq:t-T}
\end{equation}
Therefore, the scaled cumulant generating function of $A_{T}^{e}$  is  
\begin{equation}
\Lambda\left[V_{1},V_{2}\right]=\lambda\left[V_{1},V_{2}\right].\label{eq:fgc-1}
\end{equation}
We are now ready to prove that (\ref{eq:GE}) allows us to obtain the explicit forms (\ref{eq:2,5jp}) and (\ref{eq:2,5dp}).

\subsubsection{Level 2.5 for jump processes}
\label{sub:Spejp}

Using (\ref{eq:fgc-1}), relation  (\ref{eq:GE}), with $\mathit{\overrightarrow{\omega_{t}^{e}}}\equiv$$\left\{ \rho_{t}^{e},C_{t}^{e}\right\}$, becomes 
\begin{equation}
I\left[\rho,C\right]=\sup_{V_{1},V_{2}}\left\{\int_{\mathcal{E}}dx\rho(x)V_{1}(x)+\int\int_{\mathcal{E}^{2}}dxdyC(x,y)V_{2}(x,y)-\lambda\left[V_{1},V_{2}\right]\right\}.\label{eq:GEpj}
\end{equation}
The functions $V_{1}^{\star}$ and $V_{2}^{\star}$ extremizing the above expression are then obtained by solving the equations 
\begin{equation}
\left.\frac{\delta\lambda\left[V_{1},V_{2}\right]}{\delta V_{1}(x)}\right|_{V_{1}^{\star},V_{2}^{\star}}=\rho(x)\qquad\textrm{and}\qquad\left.\frac{\delta\lambda\left[V_{1},V_{2}\right]}{\delta V_{2}(x,y)}\right|_{V_{1}^{\star},V_{2}^{\star}}=C(x,y).\label{eq:fd}
\end{equation}
Furthermore, the normalization (\ref{eq:norm}) and $L_{V_{1},V_{2}}\left[r\left[V_{1},V_{2}\right]\right](x)= \lambda\left[V_{1},V_{2}\right]r\left[V_{1},V_{2}\right](x)$, lead to
\begin{equation}
\int_{\mathcal{E}}l\left[V_{1},V_{2}\right](x)L_{V_{1},V_{2}}\left[r\left[V_{1},V_{2}\right]\right](x)dx=\lambda\left[V_{1},V_{2}\right].\label{eq:defvp}
\end{equation}
From (\ref{eq:a1}), applying functional derivatives to (\ref{eq:defvp}) we obtain
\begin{equation}
\begin{cases}
l\left[V_{1},V_{2}\right](x)r\left[V_{1},V_{2}\right](x)=\frac{\delta\lambda\left[V_{1},V_{2}\right]}{\delta V_{1}(x)}\\
l\left[V_{1},V_{2}\right](x)W(x,y)\left[\exp\left(V_{2}(x,y)\right)\right]r\left[V_{1},V_{2}\right](y)=\frac{\delta\lambda\left[V_{1},V_{2}\right]}{\delta V_{2}(x,y)},\end{cases}\label{eq:fm}
\end{equation}
which, with (\ref{eq:fd}), leads to 
\begin{equation}
\begin{cases}
l\left[V_{1}^{\star},V_{2}^{\star}\right](x)r\left[V_{1}^{\star},V_{2}^{\star}\right](x)=\rho(x)\\
l\left[V_{1}^{\star},V_{2}^{\star}\right](x)W(x,y)\left[\exp\left(V_{2}^{\star}(x,y)\right)\right]r\left[V_{1}^{\star},V_{2}^{\star}\right](y)=C(x,y).\end{cases}\label{eq:fd-11}
\end{equation}
From the definitions of $l\left[V_{1},V_{2}\right]$ and $r\left[V_{1},V_{2}\right]$ as the left and right eigenvectors of $L_{V_{1},V_{2}}$,  the second equation in (\ref{eq:fd-11}) implies
\begin{equation}
\begin{cases}
\int dxC(x,y)=\left(\lambda\left[V_{1}^{\star},V_{2}^{\star}\right]+W\left[1\right](y)-V_1(y)\right)l\left[V_{1}^{\star},V_{2}^{\star}\right](y)r\left[V_{1}^{\star},V_{2}^{\star}\right](y)\\
\int dxC(y,x)=\left(\lambda\left[V_{1}^{\star},V_{2}^{\star}\right]+W\left[1\right](y)-V_1(y)\right)l\left[V_{1}^{\star},V_{2}^{\star}\right](y)r\left[V_{1}^{\star},V_{2}^{\star}\right](y),\end{cases}\label{eq:cps}
\end{equation}
where the first (second) line is obtained with an integration in $x$ ($y$). Hence, the constraint (\ref{eq:cpj}) is a necessary condition for the extremization and, moreover, using the first equation in (\ref{eq:fd-11}) we obtain  
\begin{equation}
\lambda\left[V_{1}^{\star},V_{2}^{\star}\right]+W\left[1\right](y)-V_{1}^{\star}(y)=\frac{\int dxC(x,y)}{\rho(y)}.\label{eq:AEm}
\end{equation}
Finally, from (\ref{eq:GEpj}) we obtain the rate function (\ref{eq:2,5jp}) as follows, 
\begin{align}
& I\left[\rho,C\right]  =  \int\int_{\mathcal{E}^{2}}dxdyC(x,y)V_{2}^{\star}(x,y)-\int_{\mathcal{E}}dx\rho(x)\left(\lambda\left[V_{1}^{\star},V_{2}^{\star}\right]-V_{1}^{\star}(x)\right)\nonumber\\
& =  \int\int_{\mathcal{E}^{2}}dxdyC(x,y)\ln\left[\frac{C(x,y)}{l\left[V_{1}^{\star},V_{2}^{\star}\right](x)W(x,y)r\left[V_{1}^{\star},V_{2}^{\star}\right](y)}\right]-\int_{\mathcal{E}}dx\rho(x)\left(\frac{\int dyC(y,x)}{\rho(x)}-W\left[1\right](x)\right)\nonumber\\
& =  \int\int_{\mathcal{E}^{2}}dxdyC(x,y)\ln\left[\frac{C(x,y)}{l\left[V_{1}^{\star},V_{2}^{\star}\right](x)r\left[V_{1}^{\star},V_{2}^{\star}\right](x)W(x,y)}\right]+\int\int_{\mathcal{E}^{2}}dxdyC(x,y)\ln\left[\frac{r\left[V_{1}^{\star},V_{2}^{\star}\right](x)}{r\left[V_{1}^{\star},V_{2}^{\star}\right](y)}\right]\nonumber\\
& -  \int_{\mathcal{E}}dx\rho(x)\left(\frac{\int dyC(y,x)}{\rho(x)}-\int dyW(x,y)\right)\nonumber\\
& =  \int\int_{\mathcal{E}^{2}}dxdyC(x,y)\ln\left[\frac{C(x,y)}{\rho(x)W(x,y)}\right]-\int_{\mathcal{E}}dx\rho(x)\left(\frac{\int dyC(y,x)}{\rho(x)}-\int dyW(x,y)\right)\nonumber\\
& + \int_{\mathcal{E}}dx\ln\left[r\left[V_{1}^{\star},V_{2}^{\star}\right](x)\right]\int_{\mathcal{E}}dy\left(C(x,y)-C(y,x)\right).
 \end{align}
Passing from the first to the second line we used $V_{2}^{\star}(x,y)=\ln\left[\frac{C(x,y)}{l\left[V_{1}^{\star},V_{2}^{\star}\right](x)W(x,y)r\left[V_{1}^{\star},V_{2}^{\star}\right](y)}\right]$, which follows from (\ref{eq:fd-11}),
and equation (\ref{eq:AEm}). Moreover, in the last equality we used the first equation in (\ref{eq:fd-11}) and the last term is zero due to the constraint (\ref{eq:cpj}), thus leading to expression (\ref{eq:2,5jp}) for the rate function.

\subsubsection{Level 2.5 for diffusion Processes}
\label{sub:Spedp}

Using (\ref{eq:fgc-1}), for diffusion processes (\ref{eq:GE}) becomes 
\begin{equation}
I\left[\rho,j\right]=\sup_{V_{1},V_{2}}\left\{\int_{\mathcal{E}}dx\rho(x)V_{1}(x)+j(x).V_{2}(x)-\lambda\left[V_{1},V_{2}\right]\right\}.\label{eq:GEpj-2}
\end{equation}
The following three change of variables lead to the final expression (\ref{eq:2,5dp}).	 
\begin{itemize}

\item First, $\left(V_{1},V_{2}\right)\rightarrow\left(V'_{1}=\ln\left(r\left[V_{1},V_{2}\right]\right),V_{2}\right)$, leading to
\begin{equation}
I\left[\rho,j\right]=\sup_{V'_{1},V_{2}}\left\{\int_{\mathcal{E}}dx\rho(x)\left(-\exp\left(-V'_{1}(x)\right)L_{0,V_{2}}\left[\exp\left(V'_{1}\right)\right](x)\right)+j(x).V_{2}(x)\right\}.\label{eq:int1}
\end{equation} 
This is proved  in appendix A. Note that $\ln\left(r\left[V_{1},V_{2}\right]\right)$ is well defined because $r\left[V_{1},V_{2}\right]$ is positive (from the Perron-Frobenius theorem).

\item Second, $\left(V'_{1},V_{2}\right)\rightarrow\left(V'_{1},V'_{2}=V_{2}+\nabla V'_{1}\right)$,
leading to 
\begin{align}
& I\left[\rho,j\right]  =  -\inf_{V'_{1}}\left(\int_{\mathcal{E}}dxj(x).\nabla V'_{1}\right)\nonumber\\
& -\inf_{V''_{2}}\left(\int_{\mathcal{E}}dx\left[\left(V'_{2}-\left(\rho D\right)^{-1}\left(j-J_{\rho}\right)\right)\frac{\rho D}{2}\left(V'_{2}-\left(\rho D\right)^{-1}\left(j-J_{\rho}\right)\right)\right]\right)\nonumber\\
& +\int dx\left(j-J_{\rho}\right)\frac{\left(\rho D\right)^{-1}}{2}\left(j-J_{\rho}\right).
 \label{eq:int2}
 \end{align}
This is proved  in appendix B.

\item Third, $\left(V'_{1},V'_{2}\right)\rightarrow\left(V'_{1},V''_{2}=V'_{2}-\left(\rho D\right)^{-1}\left(j-J_{\rho}\right)\right)$, finally gives
\begin{align}
& I\left[\rho,j\right]=-\inf_{V'_{1}}\left(\int_{\mathcal{E}}dxj(x).\nabla V'_{1}\right)\nonumber\\
& -\inf_{V''_{2}}\left(\int_{\mathcal{E}}dxV''_{2}(x)\frac{\rho D}{2}(x)V''_{2}(x)\right)+\int dx\left(j-J_{\rho}\right)\frac{\left(\rho D\right)^{-1}}{2}\left(j-J_{\rho}\right).\label{eq:int3}
\end{align}
The first term vanishes with fulfillment of  the constraint (\ref{eq:cpd}) and is $-\infty$ otherwise, while the second term vanishes. This last equation gives the final form (\ref{eq:2,5dp}). 
\end{itemize}

\section{Stationary Fluctuation Relation at the level 2.5}
\label{sec:ltFR}

We now consider the fluctuating entropy $\mathbb{J}_{T}$, which is obtained from the action functional (\ref{actionfunct}) setting $\mu_{0}(dx)=\mu_{0}^{b}(dx)=dx$.
We define the function 
\begin{equation}
\mathbb{J}_{T}/T=w(\rho_T^e,C_T^e)\qquad\textrm{and}\qquad \mathbb{J}_{T}/T=w(\rho_T^e,j_T^e),
\end{equation}
for pure jump and diffusion processes, respectively. From formulas (\ref{eq:Wjp}) and (\ref{eq:Wdp}), this function reads
\begin{equation}
w(\rho,C)=\int dxdyC(x,y)\ln\left[\frac{W(x,y)}{W(y,x)}\right]\qquad\textrm{and}\qquad w(\rho,j)=2\int dx\widehat{A_{0}}\left(x\right).D^{-1}\left(x\right)j(x),\label{eq:Wjp-1}
\end{equation}
The choice $F_{\left[0,T\right]}=\delta(\mathbb{\rho}^e_{T}-\rho,C^e_{T}-C)$ for pure jump and
 $F_{\left[0,T\right]}=\delta(\mathbb{\rho}^e_{T}-\rho,j^e_{T}-j)$ for diffusion processes in (\ref{fluC})  gives the finite time relation 
\begin{equation}
\begin{cases}
\mathbb{P}_{\mu_{0},L}(\rho_{T}^{e}=\rho,C_{T}^{e}=C^{t})=\exp\left(-Tw(\rho,C)\right)\mathbb{P}_{\mu_{0},L}(\rho_{T}^{e}=\rho,C_{T}^{e}=C)\\
\mathbb{P}_{\mu_{0},L}(\rho_{T}^{e}=\rho,j_{T}^{e}=-j)=\exp\left(-Tw(\rho,j)\right)\mathbb{P}_{\mu_{0},L}(\rho_{T}^{e}=\rho,j_{T}^{e}=C)\end{cases},\label{eq:crooks-1}
\end{equation}
where we used  the general relations $\rho_{T}^{e}\circ R=\rho_{T}^{e}$, $j_{T}^{e}\circ R=-j_{T}^{e}$, and $C_{T}^{e}\circ R=\left(C_{T}^{e}\right)^{t}$, with the index $t$ indicating transposition.
With the rate function for the large deviations at the level 2.5 obtained in the last section, the
large time asymptotic of both sides of the previous relation becomes the stationary fluctuation relation at level 2.5 
\begin{equation}
I(\rho,C^{t})=w\left[\rho,C\right]+I(\rho,C)\qquad\textrm{and}\qquad I(\rho,-j)=w\left[\rho,j\right]+I(\rho,j).\label{eq:gc2,5}
\end{equation}
From this relation, with the contraction $I(w)=\min_{w(\rho,C)=w}\left[I(\rho,C)\right]$ ( or $I(w)=\min_{w(\rho,j)=w}\left[I(\rho,j)\right]$ for diffusion processes), 
we obtain the stationary fluctuation relation
\begin{equation}
I(-w)=I(w)+w.\label{eq:GCb}
\end{equation}
This symmetry on the rate function of $\mathbb{J}_{T}$ is the GCEM symmetry. This relation can also be obtained from the transient fluctuation relation (\ref{eq:crooks}). 
We note that currents with such a symmetry in the rate function that are different from the fluctuating entropy $\mathbb{J}_{T}$ have been found in \cite{Bar1,Bar2,Bar3}. Investigating, the relation between this 
symmetric non-entropic currents and large deviations at the level 2.5 would be interesting. 

\begin{acknowledgements}
We thank Krzysztof Gawedzki for helping in the proof presented in section 4.2.4 and Hugo Touchette for carefully reading the manuscript. 
\end{acknowledgements}
\appendix

\section{Proof of (\ref{eq:int1}) }
We prove relation (\ref{eq:int1}) from relation (\ref{eq:GEpj-2}).
Writing
\begin{equation}
\left(L_{0,V_{2}}+V_{1}\right)r\left[V_{1},V_{2}\right](x)=\lambda\left[V_{1},V_{2}\right]r\left[V_{1},V_{2}\right](x),\label{eq:sep}
\end{equation}
we obtain
\begin{equation}
V_{1}-\lambda\left[V_{1},V_{2}\right]=-\left(r\left[V_{1},V_{2}\right](x)\right)^{-1}L_{0,V_{2}}\left(r\left[V_{1},V_{2}\right]\right)(x).\label{eq:sep2}
\end{equation}
With this last equation (\ref{eq:GEpj-2}) becomes 
\begin{eqnarray}
I\left[\rho,j\right] & = & \sup_{V_{1},V_{2}}\left(\int_{\mathcal{E}}dx\rho(x)\left(V_{1}(x)-\lambda\left[V_{1},V_{2}\right]\right)+j(x).V_{2}(x)\right)\\
&=& \sup_{V_{1},V_{2}}\left(\int_{\mathcal{E}}dx\rho(x)\left(-\left(r\left[V_{1},V_{2}\right](x)\right)^{-1}L_{0,V_{2}}\left(r\left[V_{1},V_{2}\right]\right)(x)\right)+j(x).V_{2}(x)\right)\label{eq:GEpj-2-1}\nonumber\\
 &=&\sup_{V'_{1},V_{2}}\left(\int_{\mathcal{E}}dx\rho(x)\left(-\exp\left(-V'_{1}(x)\right)L_{0,V_{2}}\left[\exp\left(V'_{1}\right)\right](x)\right)+j(x).V_{2}(x)\right),\nonumber
 \end{eqnarray}
where $V_1'= \ln r(V_1,V_2)$.

\section{Proof of (\ref{eq:int2}) }
The goal here is to prove relation  (\ref{eq:int2})  from  (\ref{eq:int1}).
From a direct calculation we obtain
\begin{equation}
\exp\left(-V'_{1}\right)L_{0,V_{2}}\left(\exp V'_{1}\right)=L_{0,V_{2}+\nabla V'_{1}}[1].\label{eq:simpt}
\end{equation}
Relation (\ref{eq:int1}) then becomes
\begin{align}
& I\left[\rho,j\right]  =  \sup_{V'_{1},V_{2}}\left(\int_{\mathcal{E}}dx\left(j(x).V_{2}(x)-\rho(x)L_{0,V_{2}+\nabla V'_{1}}[1](x)\right)\right)\nonumber\\
& = \sup_{V'_{1},V'_{2}}\left(\int_{\mathcal{E}}dx\left(-j(x).\nabla V'_{1}+j(x).V'_{2}(x)-\rho(x)L_{0,V'_{2}}[1](x)\right)\right)\nonumber\\
& =  -\inf_{V'_{1}}\left(\int_{\mathcal{E}}dxj(x).\nabla V'_{1}\right)+\sup_{V_{2}'}\left(\int_{\mathcal{E}}dx\left(j(x).V'_{2}(x)-\rho(x)L_{0,V'_{2}}[1](x)\right)\right).
\end{align}
We obtain the final relation (\ref{eq:int2}) with $L_{0,V'_{2}}[1]  =  \widehat{A_{0}}.V'_{2}+V'_{2}.\frac{D}{2}.V'_{2}+\nabla.\left(\frac{D}{2}.V'_{2}\right)$
and the algebraic manipulation 
\begin{align}
& \int_{\mathcal{E}}dx\left(j(x).V'_{2}(x)-\rho(x)L_{0,V'_{2}}[1](x)\right)  =  \int_{\mathcal{E}}dx\left(j(x).V'_{2}(x)-\rho(x)\left(\widehat{A_{0}}.V'_{2}+V'_{2}.\frac{D}{2}.V'_{2}+\nabla.\left(\frac{D}{2}.V'_{2}\right)\right)\right)\nonumber\\
& =  \int_{\mathcal{E}}dxj(x).V'_{2}(x)-\left[\rho(x)V'_{2}.\frac{D}{2}.V'_{2}+V'_{2}.\left(\widehat{A_{0}}\rho(x)-\frac{D}{2}.\nabla\rho+j\right)\right]\nonumber\\
& =  \int_{\mathcal{E}}dx\left[-\rho(x)V'_{2}.\frac{D}{2}.V'_{2}+V'_{2}.\left(j-J_{\rho}\right)\right]\nonumber\\
& =  -\int_{\mathcal{E}}dx\left[\left(V'_{2}-\left(\rho D\right)^{-1}\left(j-J_{\rho}\right)\right)\frac{\rho D}{2}\left(V'_{2}-\left(\rho D\right)^{-1}\left(j-J_{\rho}\right)\right)-\left(j-J_{\rho}\right)\frac{\left(\rho D\right)^{-1}}{2}\left(j-J_{\rho}\right)\right],
 \end{align}
which included formal integration by parts .

\end{document}